\documentclass[12pt]{article}
\usepackage[T2A]{fontenc}
\usepackage[utf8]{inputenc}
\usepackage[russian,english]{babel}
\usepackage{mathtools}
\usepackage{amsfonts,amssymb,amsmath} 
\usepackage{graphicx} 
\usepackage{cite}
\usepackage{hyperref}
\hypersetup{colorlinks=true, linkcolor=blue, filecolor=blue, urlcolor=blue} 

\usepackage{algorithm}

\newtheorem{proposition}{Proposition}

\allowdisplaybreaks

\begin{document}

\title{Krotov Type Optimization of Coherent and Incoherent Controls for Open Two-Qubit Systems\footnote{The paper was prepared for 
{\it Bulletin of Irkutsk State University, Series Mathematics}
following involvement of the authors in holding section ``Quantum Control''
of the International school-seminar ``Nonlinear Analysis and Extremal Problems''
(\url{https://conference.icc.ru/event/5/}), which was held in Irkutsk and online 
in July 2022. The section ``Quantum Control'' was organized on July 17th, 2022, 
where O.M. was the chair. Irkutsk (Matrosov Institute for System Dynamics
and Control Theory of Siberian Branch of Russian Academy of Sciences,
Irkutsk State University) is known as a place where some ideas of Krotov type methods were developed.}}

\author{ Oleg~V.~Morzhin$^{1,2,}$\footnote{E-mail: \url{morzhin.oleg@yandex.ru};~ 
    \href{http://www.mathnet.ru/eng/person30382}{mathnet.ru/eng/person30382};~ 
    \href{https://orcid.org/0000-0002-9890-1303}{ORCID 0000-0002-9890-1303}} 
    \quad and \quad 
Alexander~N.~Pechen$^{1,2,}$\footnote{E-mail: \url{apechen@gmail.com};~ 
    \href{http://www.mathnet.ru/eng/person17991}{mathnet.ru/eng/person17991};~ 
    \href{https://orcid.org/0000-0001-8290-8300}{ORCID 0000-0001-8290-8300}} 
    \vspace{0.2cm} \\
$^1$ Department of Mathematical Methods for Quantum Technologies,\\
    Steklov Mathematical Institute of Russian Academy of Sciences, \\
8 Gubkina St., Moscow, 119991, Russia;\\
$^2$ Quantum Engineering Research and Education Center, \\
    University of Science and Technology MISIS,\\
6 Leninskiy Prospekt, Moscow, 119991, Russia}

\date{}
\maketitle

\begin{abstract} This work considers two-qubit open quantum systems driven 
by coherent and incoherent controls. Incoherent control induces 
{\it time-dependent decoherence rates} via time-dependent spectral 
density of the environment which is used as a resource for controlling the system. 
The system evolves according to the Gorini--Kossakowski--Sudarshan--Lindblad 
master equation with time-dependent coefficients. For two types of interaction 
with coherent control, three types of objectives are considered: 
1)~maximizing the Hilbert--Schmidt overlap between the final 
and target density  matrices; 2)~minimizing the Hilbert--Schmidt 
distance between these matrices; 3)~steering 
the overlap to a given value. For the first problem, we develop 
the Krotov type methods directly in terms of density matrices with 
or without regularization for piecewise continuous constrained controls 
and find the cases where the methods produce (either exactly or with 
some precision) zero controls which satisfy the Pontryagin maximum principle 
and produce the overlap's values close to their upper estimates. 
For the problems 2) and 3), we find cases when the dual annealing method 
steers the objectives close to zero and produces a non-zero control.

\vspace{0.3cm}

\end{abstract} 

\rightline{\it Dedicated to the 90th anniversary of the birth and to the memory}
\rightline{\it of Vadim Fedorovich Krotov (1932 --- 2015)} 

\section{Introduction}\label{Introduction}

Optimal control of quantum systems attracts high interest due to various 
applications ranging from quantum computing to laser 
chemistry~\cite{KochEPJQuantumTechnol2022}. 
In~many situations, controlled quantum systems are open, 
i.e., interacting with the environment. While in some situations the environment 
is treated as an obstacle, in other cases it can be a useful control 
resource, as for example was proposed to do via {\it incoherent control} in  \cite{PechenPRA2006, PechenPRA2011}, 
where spectral (generally time-dependent and non-equilibrium) 
density of incoherent photons is used as control function jointly with
{\it coherent control} by lasers to manipulate the quantum system dynamics.  

In the approach of~\cite{PechenPRA2006}, which is used in the present work, the incoherent control induces \textit{time-dependent decoherence rates} $\gamma_k(t)$ via \textit{time-dependent spectral density of the environment} $n(t)$, in addition to coherent control $u$, so that density matrix evolves according to the master equation
\begin{equation}
\label{eq:ME}
\dot{\rho}(t) = -i[H_0 + H_{c(t)},\rho(t)] + \sum\nolimits_k \gamma_k(t) {\cal D}_k(\rho(t)),
\end{equation}
where $H_0$ is the free Hamiltonian, $H_{c(t)}$ is the Hamiltonian induced by control $c=(u,n)$, ${\cal D}_k$ is a Gorini--Kossakowski--Sudarshan--Lindblad (GKSL) dissipator, and $[A, B] = AB - BA$ is the commutator of matrices $A$ and~$B$. In addition to general consideration, two physical classes of the environment were exploited --- incoherent photons and quantum gas, with two explicit forms of ${\cal D}_k$ derived in the weak coupling limit (WCL) and low density limit, respectively. In~\cite{PechenPRA2011}, it was shown that for the master equation~(\ref{eq:ME}) with ${\cal D}_k$ derived in the WCL (describing atom interacting with photons) generic $N$-level quantum systems become approximately completely controllable in the set of density matrices. Following this general approach, various control problems for one- and two-qubit open systems controlled by simultaneous coherent and incoherent controls were considered~\cite{MorzhinLJM2019, LokutsievskiyJPA2021, PetruhanovPhotonics2023}.

Various tools are used in quantum optimal control such as Pontryagin maximum principle (PMP) \cite{BoscainPRXQuantum2021}, Krotov type methods \cite{Kazakov_Krotov1987, Tannor1992, Krotov2009, Morzhin_Pechen_2019, Goerz_NJP_2014_2021, Jager2014}, \cite[\S~6.5]{KrotovBook1996}), Hamilton--Jacobi--Bellmann equation~\cite{Gough2005}, Zhu---Rabitz \cite{ZhuRabitz1998} method, GRadient Ascent Pulse Engineering (GRAPE) \cite{KhanejaGRAPE2005, Jager2014, PetruhanovPhotonics2023}, GRAPE with quasi-Newton optimizers~\cite{deFouquieres2011}, 
speed gradient method~\cite{Fradkov2022}, 
gradient free Chopped RAndom Basis (CRAB) optimization \cite{CanevaPRA2011}, genetic algorithms \cite{JudsonRabitz1992, PechenPRA2006}, machine learning \cite{Dong_Chen_Tarn_Pechen_Rabitz_2008}. {\it Riemannian gradient optimization} approach over {\it complex Stiefel manifolds} for controlling open quantum systems for quantum technologies was developed in~\cite{OzaJPA2009}.  

Theory of optimal control contains the approach based on deriving special {\it nonlocal exact increment formulas} and constructing the corresponding tools for various problems (e.g., \cite{AntonikSrochko1998, SrochkoVA_Book2000, Buldaev2002, Arguchintsev2009, BuldaevMorzhinIrkutsk2009, MorzhinAutomRemoteControl2012}). For the problems linear in real-valued states (including bilinear), e.g., the $x$- and $\psi$-procedures ($\psi$ is co-state) (\cite[Ch.~1]{SrochkoVA_Book2000}, 
\cite[\S~5.4]{VasilievOV_Book1999}, \cite[pp.~15--16, 31--32]{Buldaev2002}, \cite[Sec.~3]{Arguchintsev2009}) are known, where the $x$-procedure is an adaptation of the general Krotov method (see about this method, e.g., \cite{KrotovFeldmanIzvAN1983}, \cite[Ch.~6]{KrotovBook1996}) without regularization, while the $\psi$-procedure is symmetrical one.

This article considers two-qubit ($N=4$) open quantum systems evolving according to the GKSL master equation  with simultaneous coherent and incoherent controls, with two types 
of interaction with coherent control.  Sec.~\ref{Sec2} considers three types of objectives: 1)~maximizing the overlap $\langle \rho(T), \rho_{\rm target} \rangle$; 2)~minimizing  
$\| \rho(T) - \rho_{\rm target}\|$; 3)~steering 
the overlap to a given $M \in (0,1)$, i.e. minimizing 
$\lvert\langle \rho(T), \rho_{\rm target} \rangle - M\rvert$. In Sec.~\ref{Sec3}, we develop for the problem~1) the Krotov type methods  in terms of density matrices with or without 
regularization for piecewise continuous 
constrained controls and find the cases where the methods produce 
(either exactly or with some precision) zero controls
which satisfy the PMP and produce overlap's values which, for sufficiently large $T$, are close to $\max\limits_\rho\langle \rho, \rho_{\rm target}\rangle$. In Sec.~\ref{Sec4}, for the problems 2) and 3), we use a parameterized class of controls and find cases when
the dual annealing method steers objectives close to zero and produces non-zero control.
Sec.~\ref{SecConclusions} resumes the article.

\section{Two-Qubit System and Objective Functionals} 
\label{Sec2} 

For two qubits, the Hilbert space is
$\mathcal{H} =  \mathbb{C}^2 \otimes \mathbb{C}^2$. Consider the following GKSL master equation with coherent and incoherent controls:
\begin{equation}
\dot\rho(t) = -i \big[H_0 + H_{c(t)}, \rho(t) \big] + \varepsilon \mathcal{L}^D_{n(t)}(\rho(t)), 
\quad \rho(0) = \rho_0, \quad t \in [0,T], 
\label{GKSL_system} 
\end{equation}
where $\rho(t) : \mathcal{H} \to \mathcal{H}$ is density matrix (positive semi-definite, $\rho(t) \geq 0$, with unit trace, ${\rm Tr}\rho(t) = 1$); the parameter $\varepsilon > 0$
describes strength of the coupling between the system and its environment; 
the Hamiltonian $H_{c(t)} = \varepsilon H_{{\rm eff}, n(t)} + H_{u(t)}$, where scalar coherent control $u$ and incoherent control $n=(n_1, n_2)$ are   
considered as piecewise continuous on $[0, T]$, control $c=(u,n)$; 
effective Hamiltonian $H_{{\rm eff}, n(t)}$ depends on $n(t)$ and describes the Lamb shift,
Hamiltonian $H_{u(t)}$ describes interaction with $u(t)$, $\mathcal{L}^D_{n(t)}(\rho(t))$ is the controlled superoperator
of dissipation,  $\rho_0$ is the initial density matrix. 
The system of units is such  that the Planck's constant $\hbar=1$. Consider
\begin{eqnarray*}
&&H_0 := H_{0,1} + H_{0,2}, \quad 
H_{0,j} := \frac{\omega_j}{2} W_j, \quad 
W_1 := \sigma_z \otimes \mathbb{I}_2, \quad W_2 := \mathbb{I}_2 \otimes \sigma_z, \\
&& H_{{\rm eff}, n(t)} := \sum\limits_{j=1}^2 H_{{\rm eff}, n_j(t)},
\quad H_{{\rm eff}, n_j(t)} := \Lambda_j n_j(t) W_j, \quad H_{u(t)} := V u(t),\\
&& V = V_1 := Q_1 \otimes \mathbb{I}_2 + \mathbb{I}_2 \otimes Q_2
\quad \text{or} \quad  
V = V_2 := Q_1 \otimes Q_2, \\
&&Q_j := \sum\nolimits_{\alpha = x,y,z} \lambda_{\alpha}^j \sigma_{\alpha} = \sin\theta_j \cos\varphi_j \sigma_x + \sin\theta_j \sin\varphi_j \sigma_y + \cos\theta_j \sigma_z,  
\end{eqnarray*}
where $j = 1,2$, the parameters $\omega_j,~\Lambda_j > 0$, $\lambda^j := (\lambda_x^j, \lambda_y^j, \lambda_z^j) \in \mathbb{R}^3$ is a given unit vector, $\sigma_x = \begin{pmatrix}
	0 & 1 \\
	1 & 0
\end{pmatrix}$,
$\sigma_y = \begin{pmatrix}
	0 & -i \\
	i & 0
\end{pmatrix}$, and
$\sigma_z = \begin{pmatrix}
	1 & 0 \\
	0 & -1
\end{pmatrix}$   
are Pauli matrices. The work~\cite{MorzhinPechenQIP2023} considers the case $Q_1=Q_2=\sigma_x$, or, equivalently, $\theta_j = \pi/2$ and $\varphi_j = 0$.

The superoperator of dissipation acts on $\rho(t)$ as
\begin{align*}
\mathcal{L}^D_{n(t)}(\rho(t)) :=&  \sum\nolimits_{j=1}^2 \Big[ \Omega_j \left( n_j(t) + 1 \right) \Big(2 \sigma_j^- \rho(t) \sigma_j^+ - \left\{ \sigma_j^+ \sigma_j^-, \rho(t) \right\}\Big) \nonumber \\ 
&+
\Omega_j n_j(t) \Big(2 \sigma_j^+ \rho(t) \sigma_j^-  -   \left\{ \sigma_j^- \sigma_j^+, \rho(t) \right\} \Big) \Big], 
\end{align*}
where the parameters $\Omega_j > 0$, $j=1,2$; $\{A, B\} = AB + BA$ denotes anti-commutator of matrices $A$ and $B$. The matrices $\sigma_1^{\pm} = \sigma^{\pm} \otimes \mathbb{I}_2$, $\sigma_2^{\pm} = \mathbb{I}_2 \otimes \sigma^{\pm}$ are obtained with $\sigma^+ = \begin{pmatrix}
	0 & 0 \\ 1 & 0
\end{pmatrix}$, 
$\sigma^- = \begin{pmatrix}
	0 & 1 \\ 0 & 0
\end{pmatrix}$, $\mathbb I_2$ is the $2\times 2$ identity matrix. 

Set the following constraints including the required $n_j(t)\geq 0$: 
\begin{eqnarray}
|u(t)| \leq \mu, \quad 0 \leq n_j(t) \leq n_{\max}, \quad j = 1,2, \quad \mu,~n_{\max} > 0, ~ t \in [0, T].~ 
\label{constraint_on_controls}
\end{eqnarray} 

For the system (\ref{GKSL_system}), consider the control objectives
\begin{align}
J_1(c) &:= \langle \rho(T), \rho_{\rm target} \rangle = {\rm Tr}(\rho(T) \rho_{\rm target}) \to \sup ~~ \text{(fixed}~ T>0),
\label{overlap_to_be_optimized} \\
J_2(c,T) &:= T + P \|\rho(T) - \rho_{\rm target}\| \to \inf, 
\label{steering_to_target_density_matrix} \\
J_3(c,T) &:= T + P \lvert \langle \rho(T), \rho_{\rm target} \rangle - M \rvert \to \inf, \quad M \in (0, 1),
\label{steering_overlap_to_M_obj_functional} 
\end{align}
where the Hilbert--Schmidt distance is $\| \rho - \sigma\| = \left[{\rm Tr}\left( (\rho - \sigma)^2 \right) \right]^{1/2}$,  $\rho(T)$ is the final state for given
controls $u,n$; $P>0$ is some penalty coefficient. 

For (\ref{steering_to_target_density_matrix}) and (\ref{steering_overlap_to_M_obj_functional}), consider the following parameterized class of controls (inspired by CRAB~\cite{CanevaPRA2011} and (12) from \cite{Kallush_Dann_Kosloff_2022}): 
\begin{eqnarray}
u(t) &=& \exp\left(-h_u \left(t - T/2\right)^2 \right) \sum\nolimits_{k=1}^K \left( A_k \sin(\nu_k t) + B_k \cos(\nu_k t) \right), ~~
\label{coherent_u_exp_sum_sin_cos} \\
n_j(t) &=& C_j \exp\left(-h_{n_j} \left(t - T/2\right)^2 \right), \qquad j = 1,2,
\label{incoherent_n1_n2_A_exp} 
\end{eqnarray} 
where $K$ and $\nu_1, \dots, \nu_K$ are given. The parameters $h_u$, $h_{n_1}$,  $h_{n_2}$, $A_1, \dots, A_K$, $B_1, \dots, B_K$, and $C_1,~C_2$ are bounded from above and below, where the zero lower bound for $C_j$ is required and other bounds are arbitrary.  

\section{Krotov Type Methods for Maximizing the~Overlap}
\label{Sec3}

For (\ref{overlap_to_be_optimized}), introduce the auxiliary objective  
\begin{eqnarray}
I^{\alpha}(c; c^{(k)}) := I(c) + \frac{\widehat{\alpha}}{2\alpha} \| c - s~ c^{(k)}\|_{L^2[0,T]}^2 \to \inf, \quad I(c) := b - J_1(c),~~
\label{auxiliary_objective_functional_I}
\end{eqnarray}
where $b\in\{\max\limits_\rho\langle \rho, \rho_{\rm target}\rangle,1\}$, $\alpha > 0$, $\widehat{\alpha} \in \{0, 1\}$ turns off and on the regularization; $c^{(k)} = (u^{(k)}, n_1^{(k)}, n_2^{(k)})$ is a given control, $s \in \{0,1\}$ switches between the two regularization's types. The works \cite{Krotov2009}, \cite[\S~6.5]{KrotovBook1996} for the 
Schr\"{o}dinger equation consider both presence and absence of regularization (with $s=0$ in our terms). The case $s = 0$ also reminds~\cite{Tannor1992}, where the Schr\"{o}dinger equation is considered. The case $s=1$ reminds the regularization in \cite[p.~61]{SrochkoVA_Book2000} leading to the projection improvement procedures under constrained control; also this case reminds the regularized Krotov type methods used in  \cite{Goerz_NJP_2014_2021, Jager2014} for various quantum problems. 

\subsection{Pontryagin Function and Krotov Lagrangian}  

For (\ref{auxiliary_objective_functional_I}), the  corresponding Pontryagin function is
\begin{eqnarray*}
h^{\alpha}(\chi,\rho,c) &=& \left\langle \chi, -i[H_c, \rho] + \varepsilon  \mathcal{L}^D_n(\rho) \right\rangle - \frac{\widehat{\alpha}}{2\alpha}\|c - s c^{(k)}(t)\|^2_2 = \nonumber \\
&=& \langle \mathcal{K}^c (\chi,\rho), c \rangle - \frac{\widehat{\alpha}}{2\alpha}\|c - s c^{(k)}(t)\|^2_2 + \overline{h}(\chi,\rho), 
\end{eqnarray*} 
where $\chi$ and $\rho$ are $4 \times 4$ density matrices; $c = (u, n_1, n_2) \in \mathbb{R}^3$; the  function $\mathcal{K}^c = (\mathcal{K}^u, \mathcal{K}^{n_1}, \mathcal{K}^{n_2})$ consists of $\mathcal{K}^u(\chi,\rho) = \left\langle \chi, -i[V, \rho] \right\rangle$ and 
\[\mathcal{K}^{n_j}(\chi,\rho) = \Big\langle \chi, -i[\Lambda_j W_j, \rho] + 
\varepsilon \Omega_j\Big(2 \sigma_j^- \rho \sigma_j^+ + 2\sigma_j^+ \rho \sigma_j^- - \big\{ \mathbb{I}_4, \rho \big\} \Big) \Big\rangle\] 
(the $4\times 4$ identity matrix $\mathbb{I}_4$ appears in $\mathcal{K}^{n_j}$ since $\sigma_j^+ \sigma_j^- + \sigma_j^- \sigma_j^+ = \mathbb{I}_4$), $j=1,2$; the term
\[
\overline{h}(\chi,\rho) = \Big\langle \chi, -i [H_S, \rho] + \varepsilon \sum\nolimits_{j=1}^2 \Omega_j \big(2 \sigma_j^- \rho \sigma_j^+ - \big\{ \sigma_j^+ \sigma_j^-, \rho \big\} \big) \Big\rangle.
\]

The Krotov Lagrangian for our problem is 
\begin{eqnarray*}
L^{\alpha}(c, \rho) &=& G(\rho(T)) - \int_0^T R^{\alpha}(t, \rho(t), c(t))dt, \\
G(\rho(T)) &=& b - \langle \rho(T),  \rho_{\rm target} \rangle + \langle \chi(T), \rho(T) \rangle - \langle \chi(0), \rho_0 \rangle, \\
R^{\alpha}(t,\rho,c) &=& \left\langle \chi(t), -i[H_c, \rho] + \varepsilon \mathcal{L}^D_n(\rho) \right\rangle + \langle \dot \chi(t), \rho\rangle - \frac{\widehat{\alpha}}{2\alpha}\|c - s c^{(k)}(t)\|^2_2. 
\end{eqnarray*}
The function $\chi$ is defined in Sec.~\ref{subsec33}. Note that $\int_0^T R^{\alpha}(t, \rho(t), c(t))dt$ is
\begin{align*}
&\int_0^T \left(\frac{d}{dt} \langle \chi(t), \rho(t) \rangle - \frac{\widehat{\alpha}}{2\alpha}\|c(t) - s c^{(k)}(t)\|^2_2 \right) dt = \langle \chi(t), \rho(t) \rangle \vert_0^T -\nonumber \\
&- \frac{\widehat{\alpha}}{2\alpha}\|c - s c^{(k)}\|^2_{L^2[0,T]} = G(\rho(T)) - b + \langle \rho(T), \rho_{\rm target} \rangle -\nonumber \\
&- \frac{\widehat{\alpha}}{2\alpha}\|c - s c^{(k)} \|^2_{L^2[0,T]},  L^{\alpha}(c,\rho) = b - \langle \rho(T),  \rho_{\rm target} \rangle + \frac{\widehat{\alpha}}{2\alpha}\|c - s c^{(k)} \|^2_{L^2[0,T]}.
\end{align*}

\subsection{Increment Formulas} 

Consider increment of $L^{\alpha}$ at (admissible) controls $c, c^{(k)}$:
\begin{eqnarray}
\label{incrementKrotovLagrangian}
L^{\alpha}(c, \rho) - L^{\alpha}(c^{(k)}, \rho^{(k)}) &=& G(\rho(T)) - G(\rho^{(k)}(T)) - \int_0^T (R^{\alpha}(t, \rho(t), c(t)) -\nonumber\\
&& -R^{\alpha}(t, \rho^{(k)}(t), c^{(k)}(t)))dt,
\end{eqnarray}
where the process $(c^{(k)}, \rho^{(k)})$ is either given or obtained at the previous iteration. 
Following the idea of \cite{KrotovFeldmanIzvAN1983}, 
set the conditions  
\begin{eqnarray*}
G(\rho(T)) \leq G(\rho^{(k)}(T)) \quad \text{and} \quad R^{\alpha}(t, \rho(t), c(t)) \geq R^{\alpha}(t, \rho^{(k)}(t), c^{(k)}(t))
\end{eqnarray*}
to make $L^{\alpha}(c, \rho) \leq L^{\alpha}(c^{(k)}, \rho^{(k)})$. To satisfy the inequality condition for $G$, we simply take $G(\rho(T)) = G(\rho^{(k)}(T))$ that gives 
\begin{eqnarray*}
G(\rho(T)) - G(\rho^{(k)}(T)) = \langle \chi(T) - \rho_{\rm target}, \rho(T) - \rho^{(k)}(T) \rangle = 0
\end{eqnarray*}
and, as the result, implies {\it the transversality condition} $\chi(T) = \rho_{\rm target}$. 

We can represent the integrand in (\ref{incrementKrotovLagrangian}) as $\Delta_1 R^{\alpha} + \Delta_2 R$, where 
\begin{eqnarray*} 
\Delta_1 R^{\alpha} &=& R^{\alpha}(t, \rho(t), c(t)) - R^{\alpha}(t, \rho(t), c^{(k)}(t)),\\
\Delta_2 R &=& R^{\alpha}(t, \rho(t), c^{(k)}(t)) - R^{\alpha}(t, \rho^{(k)}(t), c^{(k)}(t)).
\end{eqnarray*}
Here $\Delta_1 R^{\alpha}$ and $\Delta_2 R$ are constructed similarly to \cite{KrotovFeldmanIzvAN1983}, \cite[p.~243]{KrotovBook1996}. Then 
\begin{eqnarray*}
\Delta_2 R &=& \langle \chi(t), -i[H_{c^{(k)}(t)},  \rho(t) - \rho^{(k)}(t)] + \varepsilon (\mathcal{L}^D_{n^{(k)}(t)}(\rho(t)) \nonumber \\
&&- \mathcal{L}^D_{n^{(k)}(t)}(\rho^{(k)}(t))) \rangle 
+ \langle \dot \chi(t), \rho(t) - \rho^{(k)}(t) \rangle = \langle \rho(t) - \rho^{(k)}(t), \nonumber \\
&=&  i[H_{c^{(k)}(t)}, \chi(t)] + \dot \chi(t) \rangle + \varepsilon \langle \chi(t), \mathcal{L}^D_{n^{(k)}(t)}(\rho(t)) - \mathcal{L}^D_{n^{(k)}(t)}(\rho^{(k)}(t)) \rangle,
\nonumber
\end{eqnarray*}
where the anticommutativity property of commutator and cyclic permutation of matrices under trace are used. 
To satisfy the condition $\Delta_2 R \geq 0$, let $\Delta_2 R = 0$.
Transform the last term in $\Delta_2 R$, which is 
\begin{align*}
&\varepsilon \Big\langle \chi(t), \sum\nolimits_{j=1}^2 \Big[ \Big( \Omega_j (n_j^{(k)}(t) + 1)(2\sigma_j^- \rho(t) \sigma_j^+ 
- \{\sigma_j^+ \sigma_j^-, \rho(t) \}) + \Omega_j n_j^{(k)}(t) \times \nonumber \\ 
&\times (2\sigma_j^+ \rho(t) \sigma_j^- 
- \{ \sigma_j^- \sigma_j^+, \rho(t) \} ) \Big)  - \Big(\Omega_j (n_j^{(k)}(t) + 1) (2\sigma_j^- \rho^{(k)}(t) \sigma_j^+ -\nonumber \\
&- \{ \sigma_j^+ \sigma_j^-, \rho^{(k)}(t) \} ) + \Omega_j n_j^{(k)}(t) (2\sigma_j^+ \rho^{(k)}(t) \sigma_j^- 
- \{ \sigma_j^- \sigma_j^+, \rho^{(k)}(t) \} ) \Big) \Big] \Big\rangle
\end{align*}
to $\varepsilon \big\langle \mathcal{L}^{D,\dagger}_{n^{(k)}(t)}(\chi(t)), \rho(t) - \rho^{(k)}(t) \big\rangle$ with  
\begin{align*}
\mathcal{L}^{D,\dagger}_{n^{(k)}(t)}(\chi(t)) := \sum\nolimits_{j=1}^2 \Big[ \Omega_j \left( n^{(k)}_j(t) + 1 \right) \Big(2 \sigma_j^+ \chi(t) \sigma_j^- -  \left\{ \sigma_j^+ \sigma_j^-, \chi(t) \right\}\Big) +\nonumber \\    + \Omega_j n^{(k)}_j(t) \Big(2 \sigma_j^- \chi(t) \sigma_j^+  - \left\{ \sigma_j^- \sigma_j^+, \chi(t) \right\} \Big) \Big],
\end{align*}
here ``$\dagger$'' reflects that 
$\langle \chi(t), T_1(\rho(t)) - T_1(\rho^{(k)}(t)) \rangle = \langle T_1^{\dagger}(\chi(t)), \rho(t) - \rho^{(k)}(t) \rangle$ and 
$\langle \chi(t), T_2(\rho(t)) - T_2(\rho^{(k)}(t)) \rangle = \langle T_2^{\dagger}(\chi(t)), \rho(t) - \rho^{(k)}(t)\rangle$, 
where the operators $T_1 := 2\sigma_j^- \boldsymbol{\cdot} \sigma_j^+$, $T_2 := 2\sigma_j^+ \boldsymbol{\cdot} \sigma_j^-$ and $(\sigma_j^+)^{\top}=\sigma_j^-$, $(\sigma_j^-)^{\top}=\sigma_j^+$; note that  $\sigma_j^+ \sigma_j^-$, $\sigma_j^- \sigma_j^+$ are Hermitian. See also~\cite{Goerz_NJP_2014_2021}.  The condition $\Delta_2 R = 0$ leads to the {\it adjoint system} 
\begin{equation}
\dot \chi^{(k)}(t) = -i[H_{c^{(k)}(t)}, \chi^{(k)}(t)] - \varepsilon \mathcal{L}^{D,\dagger}_{n^{(k)}(t)}(\chi(t)),
\quad \chi^{(k)}(T) = \rho_{\rm target}.
\label{adjoint_system_rho_method}
\end{equation}
 The system is solved backward in time. Further, we see that 
\begin{align*}
\Delta_1 R^{\alpha} = h^{\alpha}(\chi^{(k)}(t), \rho(t), c(t)) - h^{\alpha}(\chi^{(k)}(t), \rho(t), c^{(k)}(t)) = \langle \mathcal{K}^c(\chi^{(k)}(t), \rho(t)), \nonumber \\ c(t) - c^{(k)}(t) \rangle - 
\frac{\widehat{\alpha}}{2\alpha}  \|c(t) - s c^{(k)}(t)\|^2_2 + \frac{\widehat{\alpha}}{2\alpha}\|c^{(k)}(t) - s c^{(k)}(t)\|^2_2. 
\end{align*}
The first increment formula (for $I^{\alpha}$) is
\begin{eqnarray}
\label{first_increment_formula}
&&I^{\alpha}(c; c^{(k)}) - I^{\alpha}(c^{(k)}; c^{(k)}) = -\int_0^T  \Big(
\langle \mathcal{K}^c(\chi^{(k)}(t), \rho(t)), c(t) - c^{(k)}(t) \rangle - \nonumber \\
&&\qquad - \frac{\widehat{\alpha}}{2\alpha} \|c(t) - s c^{(k)}(t)\|^2_2 \Big) dt - \frac{\widehat{\alpha}}{2\alpha} \int_0^T \|c^{(k)}(t) - s c^{(k)}(t)\|^2_2 dt. 
\end{eqnarray}
Important that this exact and nonlocal formula does not contain any residual. If $\widehat{\alpha}=0$, then $I^{\alpha}(c;c^{(k)}) = I(c)$ and $h(\chi, \rho, c) := h^{\alpha}(\chi, \rho, c)$.

Further, for deriving the second increment formula, 
we represent the integrand in (\ref{incrementKrotovLagrangian}) 
as the sum $\Delta_3 R + \Delta_4 R^{\alpha}$,  
where
\begin{eqnarray*}
\Delta_3 R &=& R(t, \rho(t), c(t)) - R(t, \rho^{(k)}(t), c(t)) = \\
&=& \langle i [H_{c(t)}, \chi(t)] + \dot \chi(t) + 
\varepsilon \mathcal{L}^{D,\dagger}_{n(t)}(\chi(t)), \rho(t) - \rho^{(k)}(t) \rangle, \\
\Delta_4 R^{\alpha} &=& R(t, \rho^{(k)}(t), c(t)) - R(t, \rho^{(k)}(t), c^{(k)}(t)) = \\ 
&=& h(\chi(t), \rho^{(k)}(t), c(t)) - h(\chi(t), \rho^{(k)}(t), c^{(k)}(t)) = \\ 
&=& \langle \mathcal{K}^c(\chi(t), \rho^{(k)}(t)), c(t) - c^{(k)}(t) \rangle + \frac{\widehat{\alpha}}{2\alpha}\|c(t) - s c^{(k)}(t)\|^2_2.
\nonumber 
\end{eqnarray*}  
For $\Delta_3 R = 0$ we obtain  
\begin{eqnarray}
\dot \chi(t) = -i[H_{c(t)}, \chi(t)] - \varepsilon \mathcal{L}^{D,\dagger}_{n(t)}(\chi(t)),
\qquad \chi(T) = \rho_{\rm target}.
\label{adjoint_system_chi_method}
\end{eqnarray}
The second increment formula (for $I^{\alpha}$) is 
\begin{eqnarray} 
&&I^{\alpha}(c; c^{(k)}) - I^{\alpha}(c^{(k)}; c^{(k)}) = -\int_0^T  \Big(
\langle \mathcal{K}^c(\chi(t), \rho^{(k)}(t)), c(t) - c^{(k)}(t) \rangle - \nonumber \\
&&\qquad - \frac{\widehat{\alpha}}{2\alpha} \|c(t) - s c^{(k)}(t)\|^2_2 \Big) dt - \frac{\widehat{\alpha}}{2\alpha} \int_0^T \|c^{(k)}(t) - s c^{(k)}(t)\|^2_2 dt.
\label{second_increment_formula}
\end{eqnarray} 

\subsection{Without Regularization: $\rho$- and $\chi$-methods}
\label{subsec33} 

Here based on (\ref{first_increment_formula}) for $\widehat{\alpha} = 0$, we construct the
{\it $\rho$-method}. To obtain a~control $c$ such that $I(c) - I(c^{(k)}) \leq 0$ and using~(\ref{first_increment_formula}) for $\widehat{\alpha} = 0$, set the condition $\langle \mathcal{K}^c(\chi^{(k)}(t), \rho(t)), c(t) - c^{(k)}(t) \rangle \geq 0$, 
$t \in [0, T]$, for implementation of which we introduce the following mappings:
\begin{eqnarray}
u^{\ast}(\rho, t) &=& {\rm arg}\max\limits_{|u| \leq \mu} (\mathcal{K}^u (\chi^{(k)}(t), \rho) u) = \nonumber \\
&=& \begin{cases}
-\mu, & \mathcal{K}^u (\chi^{(k)}(t), \rho) < 0, \\
\mu, & \mathcal{K}^u (\chi^{(k)}(t), \rho) > 0, \\
u_{\rm sing} \in [-\mu, \mu], & \mathcal{K}^u (\chi^{(k)}(t), \rho) = 0,
\end{cases} 
\label{u_maximization_mapping} \\
n_j^{\ast}(\rho, t) &=& {\rm arg}\max\limits_{n_j \in [0, n_{\max}]} (\mathcal{K}^{n_j}(\chi^{(k)}(t), \rho) n_j) = \nonumber \\
&=& \begin{cases}
0, & \mathcal{K}^{n_j} (\chi^{(k)}(t), \rho) < 0, \\
n_{\max}, & \mathcal{K}^{n_j} (\chi^{(k)}(t), \rho) > 0, \\
n_{j, {\rm sing}} \in [0, n_{\max}], & \mathcal{K}^{n_j} (\chi^{(k)}(t), \rho) = 0,
\end{cases} \qquad j = 1,2. \qquad 
\label{n_j_maximization_mapping} 
\end{eqnarray}
Define $n^{\ast}(\rho, t) := \big(n_1^{\ast}(\rho, t), n_2^{\ast}(\rho, t) \big)$ and $c^{\ast}(\rho, t) := \big(u^{\ast}(\rho, t), n^{\ast}(\rho, t))$.  

As the result, the system (\ref{GKSL_system}) is solved with $u^{\ast}(\rho(t),t)$ and  $n^{\ast}(\rho(t),t)$:
\begin{eqnarray}
\dot\rho(t) = -i \big[ H_{c^{\ast}(\rho(t), t)}, \rho(t) \big] + \varepsilon \mathcal{L}^D_{n^{\ast}(\rho(t), t)}(\rho(t)), 
\quad \rho(0) = \rho_0. ~~~
\label{rho_method_GKSL_system} 
\end{eqnarray}  

\begin{algorithm}
\caption{Algorithm of the $\rho$-method (no regularization)}
$k$-th ($k\ge 0$)  iteration ($\rho$-procedure) consists of the following steps. 

\quad 1) If $k=0$, then obtain $\rho^{(0)}$ from (\ref{GKSL_system}) and compute $I(c^{(0)})$; if $k>0$ then for the known $c^{(k)}$ solve the adjoint system~(\ref{adjoint_system_rho_method}) to obtain $\chi^{(k)}$.

\quad 2) Form the mappings (\ref{u_maximization_mapping}), (\ref{n_j_maximization_mapping}).

\quad 3) Solve the system (\ref{rho_method_GKSL_system})  and 
construct the set of pairs $\{( c,~\rho )\}$, where each pair consists of a solution $\rho$ 
of (\ref{rho_method_GKSL_system}) and $c(t) = c^{\ast}(\rho(t),t)$. 

\quad 4) Select the pair $(c^{(k+1)}, \rho^{(k+1)})$ which provides maximal decrease of $I(c)$; if there is no decrease then select any $(c^{(k+1)}, \rho^{(k+1)})$.

\quad 5) If the stopping criterion 
$\lvert I(c^{(k+1)}) - I(c^{(k)}) \rvert < \epsilon \ll 1$ is not satisfied, then set $k: = k+1$ and  go to step 1); otherwise stop. 

\end{algorithm}

The r.h.s. in (\ref{rho_method_GKSL_system}) is, in general,
discontinuous in $\rho$ and $t$. As known from the $x$- and $\psi$-procedures's theory, 
solution of such a system can be non-unique. The works \cite[Sec.~2]{SrochkoRussianMath2004},
\cite[Sec.~6]{KrotovBulatovBaturina2011}, \cite[App.~2]{KrotovAutomatRemoteControl2013} for the problems with scalar control and real-valued states  explain, for the situation 
when the switching function becomes zero, how to construct control via differentiation of the equality of switching function to zero. Our case may include  $\mathcal{K}^u (\chi^{(k)}(t), \rho(t)) = 0$, $\mathcal{K}^{n_1}(\chi^{(k)}(t), \rho(t)) = 0$, and $\mathcal{K}^{n_2}(\chi^{(k)}(t), \rho(t)) = 0$ whose differentiation gives the complicated expressions. 

Define the~sets $\Theta_1^u := \{ t \in [0,T]:  u^{(k+1)}(t) \neq u^{(k)}(t) \}$,
\begin{eqnarray*} 
\Theta_1^{n_j} &:=& \{ t \in [0,T]:  n_j^{(k+1)}(t) \neq n_j^{(k)}(t) \}, \\
\Theta_{2,\rho}^u &:=& \{ t \in [0,T]:  \mathcal{K}^u(\chi^{(k)}(t), \rho^{(k+1)}(t)) \neq 0 \}, \\
\Theta_{2,\rho}^{n_j} &:=& \{ t \in [0,T]:  \mathcal{K}^{n_j} (\chi^{(k)}(t), \rho^{(k+1)}(t)) \neq 0 \}, \quad j = 1,2. 
\end{eqnarray*} 

\begin{proposition}
\label{proposition1}
The $\rho$-procedure provides a resulting control $c^{(k+1)}$ such that the first integrand  in (\ref{first_increment_formula}) for $\widehat{\alpha} = 0$, $c(t) = c^{(k+1)}(t)$, $\rho(t) = \rho^{(k+1)}(t)$, $t \in [0, T]$ is non-negative and 
$I(c^{(k+1)}) \leq I(c^{(k)})$. The strong inequality $I(c^{(k+1)}) < I(c^{(k)})$ holds when at least one of the three sets $\Theta_1^u \cap \Theta_{2,\rho}^u$, $\Theta_1^{n_j} \cap \Theta_{2,\rho}^{n_j}$, $j = 1, 2$ has non-zero measure.
\end{proposition}

Further, based on the second increment formula (\ref{second_increment_formula}) for $\widehat{\alpha} = 0$, the {\it $\chi$-method} is formulated. Introduce the mappings 
\begin{eqnarray}
u^{\ast}(\chi, t) &=& {\rm arg}\max\limits_{|u| \leq \mu} (\mathcal{K}^u(\chi, \rho^{(k)}(t)) u),  
\label{u_maximization_mapping_chi_method} \\
n_j^{\ast}(\chi, t) &=& {\rm arg}\max\limits_{n_j \in [0, n_{\max}]} (\mathcal{K}^{n_j}(\chi, \rho^{(k)}(t)) n_j),
\quad j = 1,2.
\label{n_j_maximization_mapping_chi_method}
\end{eqnarray} 
Form $n^{\ast}(\chi, t) := \big(n_1^{\ast}(\chi, t), ~n_2^{\ast}(\chi, t) \big)$ and $c^{\ast}(\chi, t) := \big(u^{\ast}(\chi, t), n^{\ast}(\chi, t) \big)$. Introduce the system (\ref{adjoint_system_chi_method}) with $u^{\ast}(\chi(t),t)$ and $n^{\ast}(\chi(t),t)$: 
\begin{eqnarray}
\dot \chi(t) = -i \big[ H_{c^{\ast}(\chi(t), t)}, \chi(t) \big] - \varepsilon \mathcal{L}^{D,\dagger}_{n^{\ast}(\chi(t), t)}(\chi(t)),
\quad \chi(T) = \rho_{\rm target}. ~~
\label{chi_method_adjoint_system} 
\end{eqnarray}  

\begin{algorithm}
\caption{Algorithm of the $\chi$-method (no regularization)}
$k$-th ($k\ge 0$)  iteration ($\chi$-procedure) consists of the following steps.  

\quad 1) If $k=0$, then obtain $\rho^{(0)}$ from (\ref{GKSL_system}) and compute $I(c^{(0)})$.

\quad 2) Form the mappings (\ref{u_maximization_mapping_chi_method}), (\ref{n_j_maximization_mapping_chi_method}).

\quad 3) Solve the system (\ref{chi_method_adjoint_system})
and construct the set $\{(c, \chi)\}$, where, for each solution
$\chi$ of (\ref{chi_method_adjoint_system}), the corresponding control is $c(t) = c^{\ast}(\chi(t),t)$; construct the set of the corresponding 
pairs $\{( c,~\rho )\}$ by solving (\ref{GKSL_system}). 

\quad 4) Select the pair $(c^{(k+1)}, \rho^{(k+1)})$ which provides maximal decrease of $I(c)$; if there is no decrease then select any $(c^{(k+1)}, \rho^{(k+1)})$.

\quad 5) If the stopping criterion 
$\lvert I(c^{(k+1)}) - I(c^{(k)}) \rvert < \epsilon \ll 1$ is not satisfied, then set $k: = k+1$ and  go to the step 2); otherwise stop.  

\end{algorithm} 

Introduce the sets $\Theta_{2,\chi}^u := \{ t \in [0,T]:  \mathcal{K}^u (\chi^{(k+1)}(t), \rho^{(k)}(t)) \neq 0 \}$ and $\Theta_{2,\chi}^{n_j} := \{ t \in [0,T]: \mathcal{K}^{n_j} (\chi^{(k+1)}(t), \rho^{(k)}(t)) \neq 0 \}$, $j = 1,2$. 

\begin{proposition}
\label{proposition2}
The $\chi$-procedure provides a resulting control $c^{(k+1)}$ such that the first integrand 
in (\ref{second_increment_formula}) for $\widehat{\alpha} = 0$, $c(t) = c^{(k+1)}(t)$, $\chi(t) = \chi^{(k+1)}(t)$, $t \in [0, T]$ is non-negative and 
$I(c^{(k+1)}) \leq I(c^{(k)})$. The strong inequality $I(c^{(k+1)}) < I(c^{(k)})$ 
holds when at least one of the three
sets $\Theta_1^u \cap \Theta_{2,\chi}^u$, $\Theta_1^{n_j} \cap \Theta_{2,\chi}^{n_j}$, $j = 1, 2$ has non-zero measure.
\end{proposition}

\subsection{Regularized $\rho$- \& $\chi$-methods. Gradient Projection Method}

Based on the increment formulas  (\ref{first_increment_formula}) and (\ref{second_increment_formula}) for $\widehat{\alpha} = 1$, we briefly describe the corresponding regularized $\rho$- and $\chi$-methods, which are called as $\rho$-method-reg$(s,\alpha)$ and $\chi$-method-reg$(s,\alpha)$. In view of \cite[p.~61]{SrochkoVA_Book2000} and~(\ref{first_increment_formula}), we construct the following mappings by deriving the stationary points of the corresponding quadratic concave functions under~(\ref{constraint_on_controls}):
\begin{eqnarray*}
u^{\alpha}(\rho,t) &=& {\rm arg}\max\limits_{|u| \leq \mu}\Big( \mathcal{K}^u(\chi^{(k)}(t), \rho) u -
\frac{1}{2\alpha}(u - s u^{(k)}(t))^2 \Big) = \nonumber\\
&=& \begin{cases}
-\mu, & u_{\rm st}(\rho, t) < -\mu, \\
\mu, & u_{\rm st}(\rho, t) > \mu, \\
u_{\rm st}(\rho, t), & |u_{\rm st}(\rho, t)| \leq \mu,
\end{cases} \\
&&\text{where} \quad u_{\rm st}(\rho, t) = s u^{(k)}(t) + \alpha \mathcal{K}^u(\chi^{(k)}(t), \rho),\\
n_j^{\alpha}(\rho,t) &=& {\rm arg}\max\limits_{n_j \in [0, n_{\max}]} \Big( \mathcal{K}^{n_j}(\chi^{(k)}(t), \rho) n_j -
\frac{1}{2\alpha}(n_j - s n_j^{(k)}(t))^2 \Big) = \nonumber\\
&=& \begin{cases}
0, & n_{{\rm st},j}(\rho, t) < 0, \\
n_{\max}, & n_{{\rm st},j}(\rho, t) > n_{\max}, \\
n_{{\rm st},j}(\rho, t), & n_{{\rm st},j}(\rho, t) \in [0, n_{\max}],
\end{cases}\\
&&\text{where} \quad n_{{\rm st},j}(\rho, t) = s n_j^{(k)}(t) + \alpha \mathcal{K}^{n_j}(\chi^{(k)}(t), \rho). 
\end{eqnarray*} 
Consider $c^{\alpha}(\rho,t) = (c^{\alpha}(\rho,t), n_j^{\alpha}(\rho,t), j=1,2)$ and substitute it into (\ref{rho_method_GKSL_system}) instead of the mapping $c^{\ast}$. Note that the r.h.s. of this system is  
continuous in $\rho$ by construction. This gives the {\it $\rho$-method-reg$(s,\alpha)$}. 

Consider $\rho$-method-reg$(s,\alpha)$ in the following projection form:
\begin{eqnarray}
c^{(k+1)}(t) = c^{\alpha}(\rho^{(k+1)}(t), t) = {\rm Pr}_Q\big(s c^{(k)}(t) + \alpha \mathcal{K}^c(\chi^{(k)}(t), \underline{\rho^{(k+1)}(t)}) \big) \quad
\label{iterative_formula_of_regularized_rho_method}
\end{eqnarray}
where $Q = [-\mu, \mu] \times [0, n_{\max}]^2$ and ${\rm Pr}_Q$ is the orthogonal projection (in $\|\cdot\|_2$) which maps any point outside of $Q$ to a closest point in $Q$, and leaves unchanged points in $Q$. The function $\rho^{(k+1)}$ is the unique solution of (\ref{GKSL_system}), where $c^{\alpha}$ is taken instead of~$c$. 

Consider the function $f(u) := \mathcal{K}^u (u - s u^{(k)}(t)) - \frac{1}{2\alpha}(u - s u^{(k)}(t))^2$. Its value at the stationary point is $f(u_{st}) = \frac{\alpha}{2}(\mathcal{K}^u)^2 \geq 0$. We analyze the cases $u_{\rm st} > \mu$ and $u_{\rm st} < -\mu$ using zeros $u_{0,1} = s u^{(k)}(t) \in [-\mu, \mu]$ and $u_{0,2} = su^{(k)}(t) + 2\alpha \mathcal{K}^u$ of $f(u)$. Omitting the details, the result is $f(\mu) \geq 0$ for $u_{\rm st} > \mu$ and $f(-\mu) \geq 0$ for $u_{\rm st} < -\mu$. Similarly for $n_{{\rm st},j}$, $j=1,2$. Thus, the increment (\ref{first_increment_formula}) is $\leq 0$ for $c^{\alpha}$. Further, comparing (\ref{auxiliary_objective_functional_I}) with (\ref{first_increment_formula}) and using the basic property $\langle y - {\rm Pr}_X(y), {\rm Pr}_X(y)-x\rangle \geq 0$ of projection $\forall y \in \mathbb{R}^n$, $\forall x \in X \subset \mathbb{R}^n$, we obtain the following upper estimate (similar to (2.3) in \cite{AntonikSrochko1998}) for $k$-th iteration of $\rho$-method-reg$(s,\alpha)$ (regularized $\rho$-procedure): 
\begin{align*}
&I(c^{(k+1)}) - I(c^{(k)}) = -\int_0^T \langle \mathcal{K}^c(\chi^{(k)}(t), \rho(t)), c^{(k+1)}(t) - s c^{(k)}(t) \rangle dt = \nonumber \\
&= -\frac{1}{\alpha}\int_0^T \Big[\langle c^{(k+1)}(t) - s c^{(k)}(t), c^{(k+1)}(t) - s c^{(k)}(t) \rangle + \nonumber \\
&\qquad + \langle (s c^{(k)}(t) + \alpha \mathcal{K}^c(\chi^{(k)}(t)) - c^{(k+1)}(t), 
c^{(k+1)}(t) - s c^{(k)}(t) \rangle \Big] dt \leq \nonumber \\
&\leq -\frac{1}{\alpha}\int_0^T \|c^{(k+1)}(t) - s c^{(k)}(t) \|_2^2 dt, ~~ \alpha>0,~~ c^{(k+1)}(t) = c^{\alpha}(\rho^{(k+1)}(t),t).
\end{align*}

Taking into account the increment formula (\ref{second_increment_formula}),
we similarly construct {\it $\chi$-method-reg$(s,\alpha)$}, where we consider the mapping $c^{\alpha}(\chi,t) = (c^{\alpha}(\chi,t), n_j^{\alpha}(\chi,t), j=1,2)$, substitute it into (\ref{chi_method_adjoint_system}) instead of $c^{\ast}$.   

\begin{proposition}
\label{proposition3}
Each of the regularized $\rho$- and $\chi$- procedures for any $\alpha > 0$ gives such $c^{(k+1)}(t) = c^{\alpha}(\rho^{(k+1)}(t),t)$ that $I(c^{(k+1)}) \leq I(c^{(k)})$, and if $c^{(k+1)}(t) \neq c^{(k)}(t)$ at a subset with non-zero measure on $[0,T]$ then $I(c^{(k+1)}) < I(c^{(k)})$.
\end{proposition}

By analogy with \cite[p.~239--240]{KrotovBook1996}, for the increment (\ref{incrementKrotovLagrangian}) after differentiating the functions $G,~R$ one can obtain the gradient
\begin{eqnarray}
{\rm grad}\, I(c^{(k)})(t) = - \mathcal{K}^c(\chi^{(k)}(t), \rho^{(k)}(t)), \quad t \in [0, T].
\label{gradient_of_objective_functional_I}
\end{eqnarray}
The same  system (\ref{adjoint_system_rho_method}) 
as in $\rho$-method-reg$(s,\alpha)$ is used. For $s=1$, the formula (\ref{iterative_formula_of_regularized_rho_method}) reminds the following iterative formula of the one-step gradient projection method (GPM-1) based on (\ref{gradient_of_objective_functional_I}) (see also  \cite[p.~59]{SrochkoVA_Book2000}):  
\begin{eqnarray}
c^{(k+1)}(t) = {\rm Pr}_Q\big(c^{(k)}(t) + \alpha^{(k)} \mathcal{K}^c(\chi^{(k)}(t), \underline{\rho^{(k)}(t)}) \big), \quad t \in [0, T],
\label{iterative_formula_GPM1}
\end{eqnarray}
where $\alpha$ can be not only fixed but also chosen as dependent on $k$. The difference between (\ref{iterative_formula_of_regularized_rho_method}) and (\ref{iterative_formula_GPM1}) is that GPM-1 uses only the background, i.e. $(c^{(k)}, \rho^{(k)}, \chi^{(k)})$, while $\rho$-method-reg$(s,\alpha)$ uses both the background and the constructed control via $c^{\alpha}$. 

Consider the following functional to be minimized:
\begin{align*}
I^{\beta}(c) &= 1 - J_1(c) + \int\nolimits_0^T (\beta_1 u^2(t) + \beta_2 (n_1(t) + n_2(t)))dt, \quad \beta_1,~\beta_2 > 0. 
\end{align*}
Its gradient is
\begin{align}
	{\rm grad}\,I^{\beta}(c^{(k)})(t) &= \Big(- \mathcal{K}^u(\chi^{(k)}(t), \rho^{(k)}(t)) + 2\beta_1 u^{(k)}(t), \nonumber \\
	&\quad\quad -\mathcal{K}^{n_j}(\chi^{(k)}(t), \rho^{(k)}(t)) + \beta_2, ~~ j=1,2 \Big). 
	\label{grad_I_beta}
\end{align}
Using~\cite{AntipinDifferEqu1994}, \cite[Sec.~5]{MorzhinPechenQIP2023}, we will consider in Subsec.~\ref{Sec:Num36} GPM-2 in the form    
\begin{align}
c^{(k+1)}(t) &= {\rm Pr}_Q\big[c^{(k)}(t) - \alpha^{(k)}
{\rm grad}\, I^{\beta}(c^{(k)})(t) + \nonumber \\ 
&\quad + \theta^{(k)} (c^{(k)}(t) - c^{(k+1)}(t)) \big], \quad t \in [0, T], \quad \theta^{(k)} \in (0,1).
\label{iterative_formula_GPM2}
\end{align}

\subsection{PMP. Zero Controls}

By analogy with \cite[Ch.~4]{Butkovskiy_Samoilenko_book}, --- where, for the problem of maximizing the same overlap with respect to the von Neumann equation with coherent control, the corresponding formulation of the PMP in terms of density matrices was done, --- we, for our optimal control problem with (\ref{overlap_to_be_optimized}) and a fixed final time~$T$, formulate the corresponding PMP. 
\begin{proposition} (PMP). \label{proposition4}
For the system (\ref{GKSL_system}), where coherent and incoherent controls are in the class of piecewise continuous controls satisfying the constraints (\ref{constraint_on_controls}) for a fixed final time~$T>0$, if some admissible control $\widehat{c} = (\widehat{u}, \widehat{n}_1, \widehat{n}_2)$ is a strict local maximum point of $J_1(c)$ in the problem of maximizing $J_1$, then 
there exist solutions $\widehat{\rho}$ and $\widehat{\chi}$ of the systems (\ref{GKSL_system}), 
 (\ref{adjoint_system_rho_method}), where  $\widehat{c} = (\widehat{u}, \widehat{n}_1, \widehat{n}_2)$, such that the 
pointwise condition $\max\limits_{c \in Q} h(\widehat{\chi}(t), \widehat{\rho}(t), c) =
 h(\widehat{\chi}(t), \widehat{\rho}(t), \widehat{c}(t))$, $t \in [0, T]$ holds, where $Q := [-\mu, \mu] \times [0, n_{\max}]^2$. One has 
\begin{eqnarray}
&&\max\limits_{|u| \leq \mu} \left( \mathcal{K}^u(\widehat{\chi}(t), \widehat{\rho}(t)) u \right) =
 \mathcal{K}^u(\widehat{\chi}(t), \widehat{\rho}(t)) \widehat{u}(t), \quad t \in [0, T],
\label{PMP_max_condition_u} \\
&& \max\limits_{n_j \in [0, n_{\max}]} \left( \mathcal{K}^{n_j}(\widehat{\chi}(t), \widehat{\rho}(t)) n_j \right) =
 \mathcal{K}^{n_j}(\widehat{\chi}(t), \widehat{\rho}(t)) \widehat{n}_j(t), \quad t \in [0, T]. \qquad
\label{PMP_max_condition_n_j}
\end{eqnarray} 
\end{proposition}

Consider $\rho_0 = {\rm diag}(a_1, a_2, a_3, a_4)$ and $\rho_{\rm target} = {\rm diag}(b_1, b_2, b_3, b_4)$ with $a_j,~b_j \geq 0$ ($j = 1,2,3,4$), $\sum\nolimits_{j=1}^4 a_j = 1$, and $\sum\nolimits_{j=1}^4 b_j = 1$. One has $J_1(c) = \sum\nolimits_{j=1}^4 d_j \rho_{jj}(T)$. As in \cite{MorzhinPechenQIP2023}, we parameterize $\rho$ (using $x_j \in \mathbb{R}$, $j=\overline{1,16}$ for $\rho_{jj}$, $j=1,2,3,4$ and ${\rm Re}\rho_{ij}$, ${\rm Im}\rho_{ij}$ with $i<j$, $i,j = 1,2,3,4$) and $\chi$ (using $y_j \in \mathbb{R}$, $j=\overline{1,16}$) and obtain bilinear systems in terms of $x, y$ which correspond to~(\ref{GKSL_system}) and~(\ref{adjoint_system_rho_method}). In these terms, $\mathcal{K}^u(\chi,\rho) = \left\langle \chi, -i[V, \rho] \right\rangle$ is realificated as $\mathcal{K}^u(y,x)$ (via Wolfram Mathematica), but the derived form is too complicated to present it here. For these $\rho_0$ and $\rho_{\rm target}$, for the realifications of~(\ref{GKSL_system}) and~(\ref{adjoint_system_rho_method}) with $c = \overline{c} = 0$ the corresponding exact analytical solutions are obtained (via Wolfram Mathematica) for any $V$ including $V_1,~V_2$ with any $Q_1,~Q_2$ (since $H_{u(t)} = V u(t)$ becomes zero for any~$t$): $\overline{x}_i, \overline{y}_i$ have certain non-trivial forms if $i \in \{1, 8, 13, 16\}$, and $\overline{x}_i =  \overline{y}_i = 0$ if $i \in \overline{1,16}\setminus\{1,8,13,16\}$. $\mathcal{K}^u(y,x)$ is a linear combination of terms $x_i y_j$ with $i,j = \overline{1,16}$, $i \neq j$ such that if $i \in \{1,8,13,16\}$ then $j \not \in \{1,8,13,16\}$. Thus, substituting $\overline{x}_j(t), \overline{y}_j(t)$ instead of $x_j, y_j$ in $\mathcal{K}^u(y,x)$ gives  $\mathcal{K}^u(\overline{y}(t), \overline{x}(t)) \equiv 0$ at the whole $[0,T]$ for any admissible values of the system's parameters. The condition (\ref{PMP_max_condition_u}) is satisfied. After realification of $\mathcal{K}^{n_j}(\overline{y}(t), \overline{x}(t))$ ($j=1,2$) we set $\rho_0 = \frac{1}{4}\mathbb{I}_4$, $\rho_{\rm target} = {\rm diag}(1,0,0,0)$ and obtain  
\begin{eqnarray*}
\mathcal{K}^{n_1}(\overline{y}(t), \overline{x}(t)) &=& - e^{-2\varepsilon (\Omega_1 + \Omega_2) T} 
\left( e^{2\varepsilon \Omega_1 t} - 1 \right) \left(2 e^{2\varepsilon \Omega_2 T} - 1 \right) \varepsilon \Omega_1 \leq 0, \\
\mathcal{K}^{n_2}(\overline{y}(t), \overline{x}(t)) &=& - e^{-2\varepsilon (\Omega_1 + \Omega_2) T} 
\left( e^{2\varepsilon \Omega_2 t} - 1 \right) \left(2 e^{2\varepsilon \Omega_1 T} - 1 \right) \varepsilon \Omega_2 \leq 0, 
\end{eqnarray*}
where $\varepsilon,\Omega_1,\Omega_2 > 0$. Using~(\ref{constraint_on_controls}), we obtain that~(\ref{PMP_max_condition_n_j}) is satisfied for $j=1,2$. Thus, $c = \overline{c} = 0$ satisfies the PMP for any admissible values of the system's parameters and $T$. Similar result was obtained in~\cite{MorzhinPechenQIP2023}. 

\subsection{Numerical Results}\label{Sec:Num36}

The Python programs for numerical simulations were written by the first author using such tools as {\tt solve\_ivp} from {\tt SciPy}, etc. We consider (\ref{GKSL_system}) with $V = V_1$, $\varepsilon = 0.1$, $\omega_1 = 1$, $\omega_2 = \Omega_j = \Lambda_j = 0.5$ ($j=1,2$), $\rho_0 = \frac{1}{4}\mathbb{I}_4$. Set $\mu = 50$, $n_{\max} = 10$.  Controls are interpolated as piecewise constant with $10^4$ subintervals.

\begin{figure}[ht!]
\centering
\includegraphics[width=0.95\linewidth]{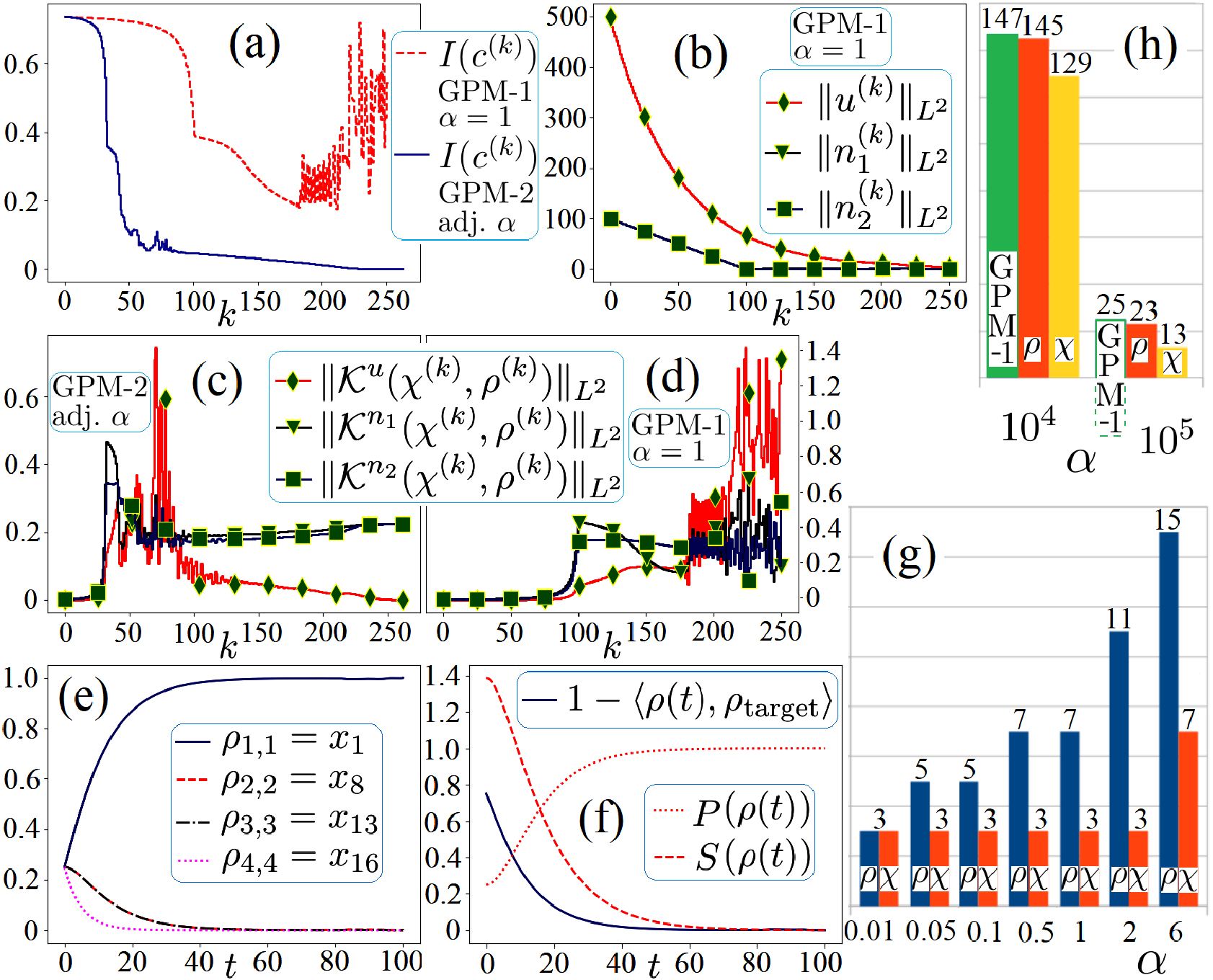}
\caption{For Case 1. Subplots~(a)--(d) show behavior vs iteration number $k$ for $c^{(0)}=(50,10,10)$ for GPM-1 and GPM-2 of: (a) objective values; (b) $L^2$ norms of controls; (c) and (d) $L^2$ norms of switching functions; (g) shows numbers of the Cauchy problems solved in $\rho$-method-reg($s=0,~\alpha$), $\chi$-method-reg($s=0,~\alpha$) for reaching $I \leq 5 \times 10^{-5}$ with the different fixed $\alpha$.
Subplot (h) for $c^{(0)}=(0, 0, 1)$ shows numbers the Cauchy problems solved in GPM-1, $\rho$-method-reg($s=1,~\alpha$), and $\chi$-method-reg($s=1,~\alpha$) for reaching $I \leq 5 \times 10^{-5}$, where $\alpha$ is either $10^4$ or $10^5$ and is the same during a~run.
Subplots (e) and (f) show for both $c^{(0)}$ $\rho_{j,j}(t)$ ($j=\overline{1,4}$), $1 - \langle \rho(t), \rho_{\rm target}\rangle$, $S(\rho(t))$, and $P(\rho(t))$ at the numerically optimized~$c$. 
}
\label{MorzhinPechenFig1} 
\end{figure} 

{\bf Case 1}. Set $\rho_{\rm target} = {\rm diag}(1,0,0,0)$ (pure quantum state), $T = 100$ and $V = V_1$ with $\theta_1 = \pi/3$, $\theta_2 = \pi/4$, $\varphi_1 = \pi/4$, $\varphi_2 = \pi/3$. 
Analytically solving (\ref{GKSL_system}) with $c=0$, we have $b - J_1 \approx 4.54 \cdot 10^{-5}$ for any $V$, where $b = 1$ is the overlap's upper bound obtained \cite{MorzhinPechenQIP2023} from $\langle \rho, \rho_{\rm target}\rangle \to \max$ and is the largest eigenvalue of $\rho_{\rm target}$. This gives a~test optimal control problem.

Set the two initial guesses: 1) $c^{(0)} = (u^{(0)}, n_1^{(0)}, n_2^{(0)}) = (\mu,n_{\max},n_{\max})$ being the furthest from $c = 0$ and giving $I(c^{(0)}) \approx 0.74$; 2)~$u^{(0)} = (0,0,1)$ being not so far from $c = 0$ and giving $I(c^{(0)}) \approx 0.33$.  

\underline{Set $c^{(0)} = (50,10,10)$}. For the functional $I^{\beta}$, we use GPM-1, GPM-2 (\ref{iterative_formula_GPM2}) with the gradient (\ref{grad_I_beta}) (here  is adapted for $I^{\beta}$). Take $\beta_1 = 0.01$, $\beta_2 = 0.1$.  Fig.~\ref{MorzhinPechenFig1}(a--d) show the following features of GPM and the dynamic control landscape of $I(c)$: 
\par 1)~we observe that up to about 80th iterations of GPM-1, $I(c^{(k)})$ varies from 0.734 to 0.689 and hence {\it changes relatively slowly} (the dashed line in Fig.~(a)) and the $L^2$-norms of $\mathcal{K}^u(\chi^{(k)}, \rho^{(k)})$, $\mathcal{K}^{n_j}(\chi^{(k)}, \rho^{(k)})$ ($j=1,2$) are {\it relatively small} (Fig.~(d)), but the $L^2$-norms of $u^{(k)}$, $n_j^{(k)}$ ($j=1,2$) {\it change significantly} (Fig.~(b)); 
\par 2)~after about 80th iterations of GPM-1 and up to $I(c^{(181)}) \approx 0.2$, the switching functions' $L^2$-norms become {\it larger} and $I$ {\it decreases significantly}; 
\par 3) after $I(c^{(181)}) \approx 0.2$, GPM-1 becomes {\it drastically worse} (Fig.~(a,d)) --- now $\alpha^{(k)}=1$ is {\it inappropriate} and GPM-1 is stopped when $k = 250$;
\par 4) GPM-2 (initially $\alpha=1$, $\theta^{(k)}=0.7$, but  $\alpha^{(k)}=0.5$, $\theta^{(k)}=0.85$, if $I \leq 0.1$, and $\alpha^{(k)}=0.3$, $\theta^{(k)}=0.9$, if $I \leq 0.05$) {\it provides $c=0$} with good precision and the stopping condition $I \leq 5 \cdot 10^{-5}$ (see the solid graph in Fig.~(a)) at the cost of 523 Cauchy problems;
\par 5) $c=0$ {\it satisfying the PMP is not a~stationary point} of $I(c)$ --- in this case, Fig.~(c) shows that $\|\mathcal{K}^u\|_{L^2}$ becomes zero while $\|\mathcal{K}^{n_1}\|_{L^2}$ and $\|\mathcal{K}^{n_2}\|_{L^2}$ do not. 

Fig.~\ref{MorzhinPechenFig1}(f) shows the von Neumann entropy $S(\rho(t)) = -{\rm Tr}(\rho(t))\log(\rho(t))$  
and purity $P(\rho(t)) = {\rm Tr}\rho^2(t)$ vs $t$ at the whole $[0,T]$. Purification of quantum states leads to $P(\rho(T)) \approx 1$, $S(\rho(T)) \approx 0$, when $T=100$, $c=0$.

For the same $c^{(0)}$, Fig.~\ref{MorzhinPechenFig1}(g) compares the numbers of the Cauchy problems for reaching $I \leq 5 \times 10^{-5}$ when we use $\rho$-method-reg($s=0, \alpha^{(k)}$) and $\chi$-method-reg($s=0, \alpha^{(k)}$) where $\alpha \in \{0.01, 0.05, 0.1, 0.5, 1, 2, 6 \}$ is the same during a~run. Both methods work sufficiently good, but the $\chi$-method is faster than the $\rho$-method for the 2nd --- 7th variants of~$\alpha$.  

In contrast to $c^{(0)} = (\mu, n_{\max}, n_{\max}) = (50, 10, 10)$, consider \underline{$c^{(0)} = (0, 0, 1)$}, because it is interesting to analyze how the methods work starting in some proximity to $c=0$. At the cost of solving only 3~Cauchy problems, $\rho$-method-reg$(s=0, \alpha=1)$ provides the zero controls and $I \leq 5 \times 10^{-5}$. The resulting $\mathcal{K}^u(y^{(0)}(t), x^{(1)}(t)) \equiv 0$, $\mathcal{K}^{n_j}(y^{(0)}(t), x^{(1)}(t)) \leq 0$, $j=1,2$ at $[0, T]$. Also at the same cost, the problem is solved via $\chi$-method-reg$(s=0, \alpha=1)$. Further, 1000 iterations as of $\rho$-method-reg$(s=1, \alpha=1)$ and of GPM-1 with the unchanged $\alpha=1$ gives only $I \approx 0.013$. 

Each of $\rho$-method-reg$(s=1, \alpha)$, 
$\chi$-method-reg$(s=1, \alpha)$, and GPM-1, where $\alpha$ is either $10^4$ or $10^5$ and is the same during a~run, provides $I \leq 5 \times 10^{-5}$. Fig.~\ref{MorzhinPechenFig1}(h) compares how many the Cauchy problems are solved in each run of these methods with the two variants for fixing~$\alpha$. The regularized $\chi$-method is essentially faster here.

Consider the $\rho$-method without regularization.
Knowing that $c=0$ satisfies the PMP, we at the first iteration, --- instead of differentiating the switching functions for deriving $u_{\rm sing}$, etc., --- try $u_{\rm sing} = 0$ in (\ref{u_maximization_mapping}), $n_{j, {\rm sing}} = 0$ in (\ref{n_j_maximization_mapping}) and obtain that $\mathcal{K}^u(y^{(0)}(t), x^{(1)}(t)) \equiv 0$, $\mathcal{K}^{n_j}(y^{(0)}(t), x^{(1)}(t)) \leq 0$, $t \in [0,T]$,  $j=1,2$ and the condition given by Proposition~1 for $I(c^{(1)}) < I(c^{(0)})$ is realized. 

{\bf Case 2}. Set $I(c) = b - J_1(c)$, $\rho_{\rm target} = {\rm diag}(0.7, 0.1, 0.1, 0.1)$, and $T = 70$, where $b = 0.7$ is the overlap's upper bound obtained \cite{MorzhinPechenQIP2023} as the largest eigenvalue of the $\rho_{\rm target}$. If $c=0$, then $I \approx 5.47 \cdot 10^{-4}$ for any $V$. Uniformly distributing $\{ \lambda^j_m \}_{m=1}^{50}$ on the unit sphere~\cite{SaffKuijlaars1997} for $\lambda_m^1 = \lambda_m^2$, the corresponding set of $Q_1,~Q_2$ is obtained. For these 50 problems, use $\rho$-method-reg$(s=0, \alpha=1)$, $c^{(0)} = (0.1, 1, 1)$. The condition $I \leq \varepsilon_{\rm stop} = 5.6 \cdot 10^{-4}$ is reached for each problem, but numbers of the solved Cauchy problems vary from 3 to~23.  
 
\section{Steering States and the Overlap: Numerical Results}
\label{Sec4} 

Consider the system (\ref{GKSL_system}) and $J_2, J_3$ to be minimized (see (\ref{steering_to_target_density_matrix}), (\ref{steering_overlap_to_M_obj_functional})). Using the parameterized class of controls (\ref{coherent_u_exp_sum_sin_cos}), (\ref{incoherent_n1_n2_A_exp}) leads to the problems of minimizing the corresponding objective functions of $T$ and ${\bf a} = (h_u, A_1, \dots, A_K, B_1, \dots, B_K, C_1, C_2, h_{n_1}, h_{n_2})$. {\tt SciPy} implementation of the dual annealing method (DAM) is used. DAM realizes some zeroth-order and stochastic approach for finite-dimensional global optimization. DAM starts from some automatically generated point. In view of the stochastic nature of DAM, we perform more than one trial for each problem.

\begin{figure}[ht]
\centering
\includegraphics[width=\linewidth]{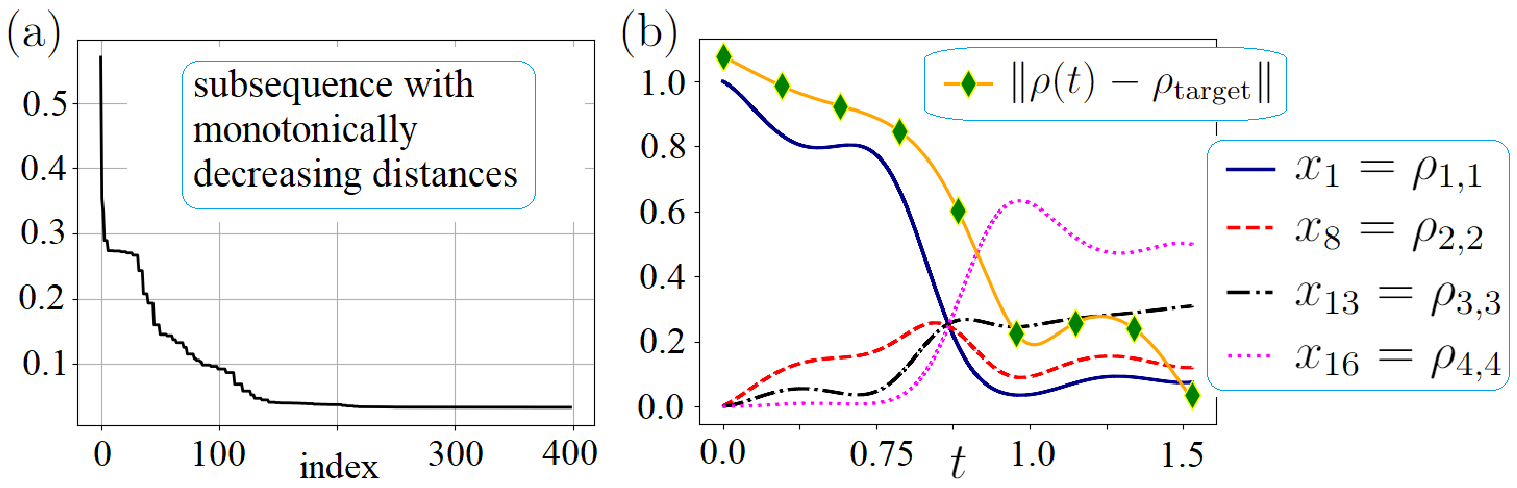} 
\caption{Steering states to $\rho_{\rm target} = {\rm diag}(0.1, 0.1, 0.3, 0.5)$ with $V=V_1$. 10~trials of DAM give non-trivial $u, n_1, n_2$. From a trial with $\| \rho(T) - \rho_{\rm target}\| \approx 0.03$: (a)~subsequence with monotonically decreasing values of $\|\rho(T) - \rho_{\rm target}\|$ during the work of DAM; (b)~the resulting $\rho_{j,j}$ and $\| \rho(t) - \rho_{\rm target}\|$ vs $t$.}
\label{MorzhinPechenFig2}
\end{figure}

{\bf Case 1}. Consider the problem of minimizing $J_2(c,T)$. Set $V = V_1$, where $\theta_1 = \theta_2 = \pi/2$, $\varphi_1 = \varphi_2 = 0$
(as considered 
in \cite{MorzhinPechenQIP2023}). Also set $\rho_0 = {\rm diag}(1,0,0,0)$ corresponding to $x_0 = (1, \text{\rm fifteen zeros})$ which is, as it was shown in \cite{MorzhinPechenQIP2023},  
singular, equilibrium state for the system $\dot x(t) = A x(t)$, i.e. for any $T$ the system's trajectory cannot leave $x_0$ without controls. Set $\rho_{\rm target} = {\rm diag}(0.1, 0.1, 0.3, 0.5)$. For (\ref{coherent_u_exp_sum_sin_cos}), (\ref{incoherent_n1_n2_A_exp}), consider 
$K=3$, $\nu_1 = 0.5$, $\nu_2 = 1$, $\nu_3 = 2$,
$h_u \in [0,2]$, $|A_j| \leq 10$, $|B_j| \leq 10$ ($j=1,2,3$), $C_j \in [0, 5]$ ($j=1,2$), $h_{n_j} \in [0,2]$ ($j=1,2$), $T \in [0.5, 2]$. After 10 trials of DAM applied to (\ref{steering_to_target_density_matrix}) with $P = 10^3$, we stop with $\|\rho(T) - \rho_{\rm target}\| \approx 0.03$ and see that, at each of these trials, the computed controls $u,~n_1,~n_2$ are not trivial. These 10 trials give such $n_1,~n_2$ that the computed values of $C_1$ in (\ref{incoherent_n1_n2_A_exp}) are from $\approx 0.08$ to $\approx 2.09$ and $C_2$ in (\ref{incoherent_n1_n2_A_exp}) are from $\approx 3.57$ to 5 under the constraints $C_j \in [0, 5]$, $j=1,2$.
Fig.~\ref{MorzhinPechenFig2} visualizes the result of some trial, where the initial distance $\| \rho(T) - \rho_{\rm target} \| \approx 0.57$, the largest occurred value is near~1. The resulting value $\| \rho(T) - \rho_{\rm target} \| \approx 0.03$ is much closer to zero. Thus, using the class of controls (\ref{coherent_u_exp_sum_sin_cos}), (\ref{incoherent_n1_n2_A_exp}) and DAM is helpful.
 
For $V = V_2$ with the same other settings, the best result from 10 trials of DAM also is $\|\rho(T) - \rho_{\rm target}\| \approx 0.03$, and there are such the trials with $\|\rho(T) - \rho_{\rm target}\| \approx 0.03$ that either $n_1$ is zero or not with non-trivial $n_2$. 

{\bf Case 2}. Consider the problem of minimizing $J_3(c,T)$. Set $V = V_1$, where $\theta_1 = \theta_2 = \pi/2$, $\varphi_1 = \varphi_2 = 0$. For the same other settings, after 10 trials of DAM, the best result for $\lvert \langle \rho(T), \rho_{\rm target} \rangle - M \rvert$ is zero. All these 10 trials give that $n_2$ is zero with various precision. 
  
\section{Conclusions} 
\label{SecConclusions} 

This article considers two-qubit open quantum systems 
driven by coherent  and incoherent controls, where the latter uses the environment 
as a resource and induces {\it time-dependent decoherence rates} so that the system evolves according to a GKSL master equation with time-dependent
coefficients. For two types of interaction with coherent control, three types of objectives 
are considered: maximizing the overlap $\langle \rho(T), \rho_{\rm target} \rangle$, 
minimizing $\| \rho(T) - \rho_{\rm target}\|$, and steering the overlap to a given value. 
For the first case we, based on \cite{KrotovBook1996, SrochkoVA_Book2000}, etc., 
formulate in terms of density matrices 
the Krotov type methods based on the nonlocal exact increment formulas 
with or without regularization for piecewise continuous 
constrained controls, and also give PMP and GPM, find the cases where the methods produce 
(either exactly or with some precision) zero controls
which satisfy the PMP and produce objectives close to $\max\limits_{\rho} \langle \rho, \rho_{\rm target} \rangle$ so that the system's 
free evolution itself achieves the goal. 
The Propositions~\ref{proposition1}--\ref{proposition3} 
state that the Krotov type procedures do not produce worse controls, 
and provide conditions when the procedures give better controls. For the other problems, we use a parameterized class of controls and find cases 
when the dual annealing method steers objectives close to zero and produces non-zero control.

Comparing with the results beyond quantum control which are the basis for us (e.g., \cite{KrotovBook1996, SrochkoVA_Book2000, VasilievOV_Book1999}), we develop the nonlocal improvement methods with or without regularization in terms of the GKSL master equation and density matrices. The work \cite{Goerz_NJP_2014_2021} develops such the Krotov type method in terms of the similar Markovian master equation and similar to $\rho$-method-reg$(s,\alpha)$, but only with coherent control and regularization. The difference from \cite{Tannor1992, Jager2014} is that these works consider for the Schr\"{o}dinger (linear in $\psi$ and control) and Gross--Pitaevskii (not linear in $\psi$) equations the corresponding Krotov type methods with regularization (with $s=0$ or $s=1$ in our terms). The works \cite[\S~6.5]{KrotovBook1996}, \cite{Krotov2009} consider the Schr\"{o}dinger equation and the improvement methods with or without regularization (with $s=0$ in our terms). Thus, we take into account and combine various known ideas and constructions to reflect the specifics (density matrices, two types of controls, two qubits, etc.) of the quantum problem and to develop the methodology of the nonlocal improvement methods to open quantum systems. Sec.~3.6 describes some interesting results of the formulated methods. Various aspects of solving the appearing equations whose r.h.s. are, in general, discontinuous in $\rho$, $\chi$, and  convergence of the methods should be carefully studied with numerical experiments in subsequent works.  
Using the dual annealing method, in Sec.~\ref{Sec4} non-trivial incoherent controls were found for some objective functionals showing the appearance of the environment as a useful resource. 

\vspace{0.5cm}

\noindent {\bf Acknowledgements.} 
Work for subsections 3.1--3.4,~3.6 was supported by the Russian Science Foundation
grant No.~22-11-00330 (\url{https://rscf.ru/en/project/22-11-00330/}), 
and performed in Steklov Mathematical Institute of Russian Academy of Sciences,
and for section~4 by the federal academic leadership program ``Priority 2030'' in MISIS.

\end{document}